\documentclass[12pt,a4paper]{article}
\usepackage{a4wide}
\usepackage{latexsym}
\usepackage{epsf}
\usepackage{amssymb}
\begin{document}
\def\be{\begin{equation}}
\def\ee{\end{equation}}
\def\bea{\begin{eqnarray}}
\def\eea{\end{eqnarray}}

\def\pd{\partial}
\def\a{\alpha}
\def\b{\beta}
\def\g{\gamma}
\def\d{\delta}
\def\m{\mu}
\def\n{\nu}
\def\t{\tau} 
\def\l{\lambda}
\def\s{\sigma}
\def\e{\epsilon}
\def\scri{\mathcal{J}}
\def\cM{\mathcal{M}}
\def\tcM{\tilde{\mathcal{M}}}
\def\RR{\mathbb{R}}
\def\CC{\mathbb{C}}

\hyphenation{re-pa-ra-me-tri-za-tion}
\hyphenation{trans-for-ma-tions}


\begin{flushright}
IFT-UAM/CSIC-99-27\\
hep-th/9907158\\
\end{flushright}

\vspace{1cm}

\begin{center}

{\bf\Large The Confining String from the Soft Dilaton Theorem}

\vspace{.5cm}

{\bf Enrique \'Alvarez}${}^{\diamondsuit,\clubsuit}$
\footnote{E-mail: {\tt enrique.alvarez@uam.es,@cern.ch}}
{\bf and C\'esar G\'omez}${}^{\diamondsuit,\spadesuit}$
\footnote{E-mail: {\tt iffgomez@roca.csic.es}} \\
\vspace{.3cm}

\vskip 0.4cm

${}^{\diamondsuit}$\ {\it  Theory Division, CERN,1211 Geneva 23, 
Switzerland \\and\\ 
Instituto de F\'{\i}sica Te\'orica, C-XVI,
  Universidad Aut\'onoma de Madrid \\
  E-28049-Madrid, Spain}\footnote{Unidad de Investigaci\'on Asociada
  al Centro de F\'{\i}sica Miguel Catal\'an (C.S.I.C.)}

\vskip 0.2cm

${}^{\clubsuit}$\ {\it Departamento de F\'{\i}sica Te\'orica, C-XI,
  Universidad Aut\'onoma de Madrid \\
  E-28049-Madrid, Spain }

\vskip 0.2cm

${}^{\spadesuit}$\ {\it I.M.A.F.F., C.S.I.C., Calle de Serrano 113\\ 
E-28006-Madrid, Spain}

\vskip 1cm


{\bf Abstract}

\end{center}

\begin{quote}
A candidate for the confining string of gauge
 theories is constructed via a representation of the 
 ultraviolet divergences of quantum 
field theory by a closed string dilaton insertion, computed
through the soft dilaton theorem.  
  The resulting (critical) confining string is conformally 
invariant, singles out naturally  $d=4$ dimensions, 
and can not be used to represent theories with Landau poles.
\end{quote}


\newpage

\setcounter{page}{1} \setcounter{footnote}{1}
The simplest way to address the problem of a string representation of a non
abelian gauge theory \cite{polyakov} consist in looking for a 
confining string background,
determined by vanishing $\sigma$ -model beta functions, such that the condition
of dilaton beta function equal to zero could be reinterpreted as the standard 
renormalization group equation:
\be
\mu\frac{dg}{d\mu}=\beta(g)
\ee
for the Yang Mills running coupling constant. This program
implies a deep geometrization of quantum field theory renormalization program,
pointing out to gravity  as the underlying dynamics 
controling the running of coupling constants \footnote{ Previous work on 
the relationship between
the renormalization group and the holographic principle include 
\cite{susskind},\cite{akhmedov},\cite{alvarez} ,\cite{alvarez1} and 
\cite{balasubramanian}.}.The next and more difficult
step, in the confining string program, ends up stablishing the precise 
relation between gauge singlets and the confining closed string spectrum.
\par
Back to the early days of string theory \cite{neveu},\cite{shapiro},
\cite{ademollo}the problem of infinities of
one loop bosonic string amplitudes, was treated in perfect analogy with the
quantum field theory renormalization program. An important subproduct
of that analysis, controling
the structure of string {\em renormalizability},
was the so called soft dilaton theorem.
\par
Let us consider a one loop open bosonic string amplitude
for N external gluons\\ 
$A^{1}(p_{1},p_{2},\ldots ,p_{N})$. On the moduli space
of the cylinder this amplitude becomes singular when the
length of the cylinder $t^{-1}\rightarrow\infty$  corresponding
to an anulus with the size of the internal hole going to
zero. In string theory the singularity of this
amplitude is exactly given by the tree level amplitude 
of emission of a soft dilaton of momentum $k=0$ that is
subsequently absorbed by the vacuum:
   
\be\label{Sing}
Sing~ ( A^{1}(p_{1},p_{2},...p_{N})=lim _{k\rightarrow 0}
J~\Delta(k)~A^{0}(p_{1},p_{2},...p_{N};k)
\ee
with $J$ the dilaton-vacuum amplitude, $\Delta(k)$ the dilaton
propagator and $A^{0}(p_{1},p_{2},...p_{N};k)$ the tree level
amplitude for $N$ gluons and one dilaton of momentum $k$.
\par
The soft dilaton theorem \cite{ademollo} precisely states that:
\be\label{sdt}
 lim_{k\rightarrow 0} A^{0}(p_{1},p_{2},...p_{N};k)= \pi g l_s^{\frac{d-2}{2}}
[l_s\frac{\pd}{\pd l_s}-
\frac{d-2}{2}g\frac{\pd}{\pd g}] A(p_1\ldots p_N)
\ee
with $g$ the open string coupling constant, and $\a'\equiv l_s^2$, 
thus implying that the singularity
in equation (\ref{Sing}) can be absorbed in a double {\em renormalization}
of $\alpha'$ and $g$.
\par
As it is clear from equation (\ref{Sing}), the singularity of 
$A^{1}(p_{1},p_{2},...p_{N})$ comes from the dilaton propagator 
$\Delta(k)$ at $k=0$. We have schematically represented in the enclosed 
figure the pertinent string diagrams. 
\begin{figure}[!ht] 
\begin{center} 
  \leavevmode \epsfxsize= 10cm \epsffile{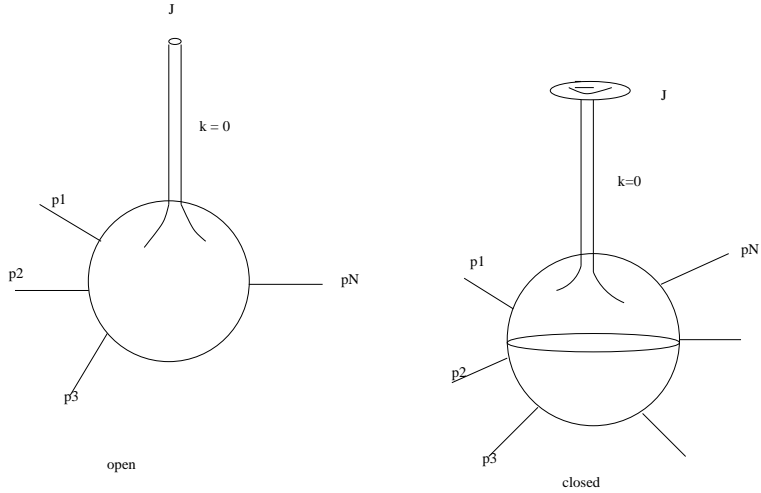}
\caption{ Topology of the sigularities in the open and closed string diagrams }
\label{fig:zero} 
\end{center} 
\end{figure}

Using the integral representation of $\Delta(k)$ and introducing
a cutoff $\mu $ as:
\be\label{delta}
\Delta_{\mu}(k)|_{k=0}=\int_{1/\mu}^{1}\frac{dz}{z}
\ee
(where $\mu$ should be interpreted as a cutoff on the {\em size} of the
long tube in Figure 1) \footnote{ It should be stressed that in ref. 
\cite{neveu} this divergence was analytically regularized, by writing
it as $\int_{0}^{1}\frac{dz}{z ^{1+\lambda}}= -\frac{1}{\lambda}$. What we
have done in order to compare with quantum field theory divergences is to
trade this linear divergence by a logarithmic one.} , 
the renormalized string tension is given by:
\be\label{renor}
\frac{1}{\a'_R(\mu)}= \frac{Jg^2}{\a'_0}\int_{1/\mu}^1 \frac{dz}{z} +
 \frac{1}{\a'_0}
\ee
From the soft dilaton theorem (\ref{sdt}) and equation (\ref{renor})
we get a {\em 
running}
string coupling constant \footnote{In references \cite{neveu},
\cite{shapiro}, the renormalization of the
string coupling constant, $g$, was interpreted as a consequence of the
tachyon singularity in the one-loop open string amplitude, in the
$t\rightarrow 0 $ limit. In \cite{ademollo}, however,the appearance of $g$
in the soft dilaton theorem was related to the transversality of
the external dilaton. We believe that the potential closed string tachyon
interpretation  of the renormalization of $g$ in 
parallel to the
transversality condition for the dilaton, could benefit from 
better understanding (cf. \cite{polyakov1},\cite{klebanov}).}
$g(\mu)$ with beta function  
\footnote{Gotten through the relationship $g_{YM}^{0} = Z^{(d-2)/8} g_{YM}$,
with $Z\equiv e^\frac{2\pi}{4} g^0 log (1/\mu )$ (cf. \cite{ademollo}).}
given by:
\be\label{beta}
\beta(g)=-g^{3}\frac{(d-2)\pi}{8}
\ee

It is interesting to observe that equation (\ref{beta}) 
for $d>2$ is of asymptotically free type.
\par
The modern approach, after Fischler-Susskind (\cite{fischler}), 
to the string tension
renormalization (\ref{renor}) stems from identifying 
$\log\mu$ with the world sheet Weyl factor 
(the Liouville field $\phi$) and to look for a background 
redefinition with a new Ricci tensor determined by:
\be
\frac{\delta(\frac{1}{\alpha'_{R}(\mu)})\eta_{\mu\nu}}{\delta log\mu}
\ee
in such a way that the $\sigma$-model beta function equations for
the string are satisfied.
\par
After the discovery of $D$-p-branes it is perfectly
natural to interpret the one loop open string amplitude 
$A^{1}(p_{1},p_{2},....p_{N})$ as a world volume amplitude
with $p_{1},p_{2},..p_{N}$ momenta in $p+1$ dimensions. In this
case the singularity in the one loop open string amplitude
can be related to a soft dilaton insertion with momentum $k$
in the {\em bulk} transversal directions. This D-brane picture
naturally yields the interpretation of  $\mu$ as a transversal coordinate.
In this set up the renormalization of $\alpha$ can be interpreted as defining
a  background metric of the type:
\be\label{metric}
ds^{2}=\frac{1}{\alpha'(\mu)}d\vec{x}_{p+1}^{2}+d\phi^{2}+d\vec{y}_{26-p-2}^2
\ee
(where $d\vec{x}_{d}^2$ is the flat metric in $\mathbb{R}^d$)
for $\phi=log \mu$, the world sheet Weyl factor or 
Liouville field (cf.\cite{
polyakov}). (The general idea of using strings with variable tension is due
to Polyakov (cf. \cite{polyakov1})).
 The condition of world sheet conformal invariance for
the background (\ref{metric}) 
ought to be consistent with the dilaton dependence $\Phi(\mu)$
dictated by soft dilaton theorem (\ref{sdt}) i.e by the
renormalized string coupling. As we will see in a moment
this is in fact the case, but only if $p=3$!
\par
Before going into more details let us highlight
the main point of this note. From the Dirac-Born-Infeld action
for a $D-3$ brane we get the standard relation between Yang Mills
coupling constant and the closed string coupling constant $g^{2}$:
\be
g_{YM}^{2}=g^{2}
\ee
where in addition $g^{2}$ is given by the vacuum expectation
value of the dilaton field.
The confining string interpretation of four dimensional
pure Yang Mills theory will be based on interpreting the
Yang Mills beta function as a {\em stringy} beta function 
governing the string one loop renormalization of $g$ as due
to soft dilaton insertions. Once we stablish this 
identification we use the soft dilaton theorem to find the associated 
running string tension $\alpha'(\mu)$. The confining string background
metric will be finally defined by a metric as (\ref{metric}).As a 
consistency check of the whole procedure we will show that
the $\sigma$-model beta functions vanish for the so defined
confining string background.
\par
It is important to stress that in this approach we read
the confining string geometry directly from the perturbative
ultraviolet behavior of pure Yang Mills gauge theory. The
identification of the string beta function,
governing the {\em renormalization}
of the string coupling constant, with the pure gauge
beta function should be understood as a generalization
to quantum field theory of expression (\ref{Sing}). In fact
what we are doing is reinterpreting the standard
ultraviolet singularities of a quantum field theory in
terms of a soft dilaton insertion in the corresponding
confining string. More precisely what we understand
as a confining string is the string interpretation, in
terms of a dilaton propagator (in bulk direction) and a dilaton 
tadpole, of the standard ultraviolet singularities
of non abelian gauge theories. In this sense we can
think of strings as a physical model of quantum field theory
renormalizability. The $D$-brane description (\cite{polchinski})
of gauge theories
will be specially suited for our purposes.
\par
The DBI action for a $D$-p-brane:
\be
S_{DBI} = T_p \int d^{p+1}\xi e^{-\Phi}
\sqrt{det[g_{ab}+b_{ab}+2\pi\a' F_{ab}]}
\ee
through an expansion in $\alpha'$, conveys the relation 
between Yang Mills bare coupling and the dilaton vacuum
expectation value:
\be\label{g}
\frac{1}{g_0^2}=T_p e^{-\Phi_0} (2\pi\alpha')^2
\ee
The one loop 
corrections to the Yang Mills coupling are given in the background field
method by:
\be\label{bf}
 (\frac{1}{g_0^2} + \frac{1}{2}C_{G,1} - C_{G,0})(F^2 + higher~derivatives)
\ee
where the coefficients $C_{G,i}$ represent the contribution
of the determinants of the gluon $(i=1)$ and ghosts $(i=0)$
quadratic terms:
\be
C_{G,i}= c_i \log (M^2/k^2)
\ee
where the scale $M$ can be appropiately chosen to cancel the
bare coupling dependence in (\ref{bf}).
\par
The simplest possible matching between equations (\ref{g}) and
(\ref{bf}) suggests a renormalization group
dependence of the dilaton field such as:
\be\label{dilaton}
T_p e^{-\Phi(k^2)}(2\pi\a')^2 F^2
\ee
This dependence will be interpreted as providing a running
string coupling constant $g(\mu^2=k^2)$ that , according to
our previous discussion will be identified with the running
string coupling constant one would obtain in the string
renormalization of string one loop infinities satisfying equation
(\ref{Sing}). In order to put this identification on more solid
grounds we need to match the renormalization group scale 
in quantum field theory with the cutoff used in 
equation (\ref{delta}) for the dilaton propagator. In order to do that
we will again profit from de $D$-brane description.
\par
To be concrete we will consider the two gluon
amplitude  corresponding to the process depicted in figure:
\begin{figure}[!ht] 
\begin{center} 
  \leavevmode \epsfxsize= 10cm \epsffile{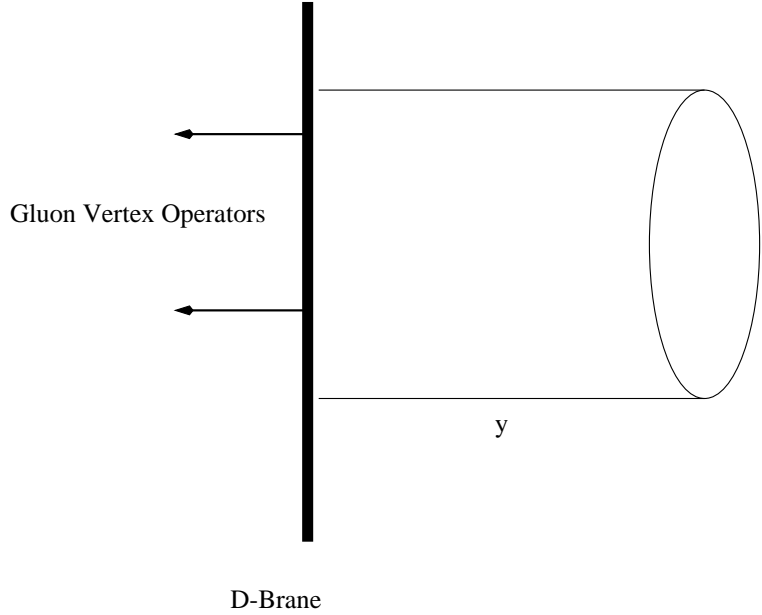}
\caption{ The planar two-gluon annulus amplitude }
\label{fig:uno} 
\end{center} 
\end{figure}

\be\label{D}
A\sim \a'^{1-d/2}\epsilon_1.\epsilon_2\frac{g^2}{(4\pi)^{d/2}}
\int dt 
\int_0^t d\nu~ t^{- d/2} 
\prod_{n=1}^{\infty} (1 - e^{-2nt})^{2-D} e^{2\a' p_1.p_2 G(\nu)} 
\pd^2 G(\nu) e^{- y^2 t /\a'}
\ee
where $G(\nu)$ is the cylinder Green function, $\nu$ is the relative position
of the vertex operators, characterized by polarizations and
momenta $\e_1,p_1$ and $\e_2,p_2$. The D-brane effect is conveyed 
by the exponential factor depending on $y$, the distance between branes.
\par
Let us now consider the limit:
\be\label{limit1}
\alpha'=0;t^{-1}=0;y=0
\ee
with
\be\label{limit2}
\alpha't=\bar{t};\frac{y}{\alpha'}=u
\ee
In equation (\ref{D}) $d=p+1$ for $p$ the dimension of
$D$-brane world volume space and $D$ represent the
dimension of the target space-time.In the limit (\ref{limit1}), 
(\ref{limit2}) amplitude (\ref{D}) for $p_{1}$ and $p_{2}$ on
shell becomes (\cite{divecchia}):
\be\label{A}
A\sim \frac{g^2}{(4\pi)^{d/2}}\epsilon_1.\epsilon_2
\int\frac{d\bar{t}}{\bar{\t}}\int_0^1 d\hat{\nu}\bar{\t}^{1-d/2}
e^{- u^2 \bar{\t}}[(D-2)(1- 2 \bar{\nu})^2 - 8]
\ee
where we have used the relationship for the Neumann Green's 
functions:
\be\label{lupa}
[e^{2\bar{\tau}} + (D-2)]\pd^2 G \sim (D-2) (1 - 2 \hat{\nu})^2 - 8
\ee
From equation (\ref{A}) we easily get:
\be
A\sim -\frac{g^2}{(4\pi)^{d/2}}\epsilon_1.\epsilon_2
\Gamma(1- d/2) (u^2)^{d/2 -1} [(D-2) - 4 (D-2) B(2,2) - 8]
\ee
(where $B(u,v)$ is Euler's Beta function),
which by standard dimensional regularization leads for $d=4$ to:
\be\label{u}
A\sim -\frac{g^2}{(4\pi)^{d/2}}\epsilon_1.\epsilon_2
u^2 ~log(u^2/\Lambda^2)[(D-2) - 4(D-2) B(2,2) - 8]
\ee
From equation (\ref{u}) it is clear that variable $u$
as defined in (\ref{limit2}) is playing the 
r\^ole of the renormalization group variable. On the other hand
$u$ as it is plain from definition (\ref{limit2}) is
nothing else but the near horizon transversal variable
introduced in reference \cite{maldacena} possessing the
natural meaning of energy in transversal bulk direction
which is precisely the meaning of the cutoff variable $\mu^2$
introduced in equation (\ref{delta}).
Using equation (\ref{dilaton}) the corresponding dilaton
field defined by (\ref{u}) will be given by:
\be\label{Phi}
e^{-\Phi(u^2)}= - [(D-2)-4(D-2)B(2,2) - 8] log(u^2/\Lambda^2)
\ee
Following our strategy what we will do now is to
identify the running of the dilaton (\ref{Phi})
\footnote{It is worth remarking that for $D=4$ the equation (\ref{u})
gives directly the well known precise factor for Yang Mills ($11/3$).
In $D=26$ instead we get zero (the correct value for $n_s=22$ 
adjoint scalars). In what follows we will stick to $D=4$.}
 with
the string renormalization of the string coupling constant due to soft
dilaton tadpole. Next and using the soft dilaton theorem
we induce from that a running dependence of world volume
metric on $u$. Finally we chek the consistency of the so defined
confining string by solving the $\sigma$-model beta functions.
Going through these steps the confining string background metric 
\footnote{With euclidean signature.}we get
is:
\be\label{metrika}
ds^2=log(u)d\vec{x}_4^2+\frac{du^2}{u^2}+ d\vec{y}_{21}^2
\ee
with a dilaton field of the type:
\be\label{dilatonio}
\Phi(u)= - log~log(u)
\ee
The $\sigma$-model beta equations are:
\be\label{uno}
R_{\mu\nu}-\nabla_{\mu}\nabla_{\nu}\Phi=0
\ee
and
\be\label{dos}
\nabla^{2}\Phi + (\nabla\Phi)^{2} = c_{extra}
\ee
It is known in general (cf, for example, \cite{callan},\cite{curci}) 
that consistency
of the equations imply that $c_{extra}$ must be constant 
(equal to $\frac{2(D-10)}{3}$ in the (world-sheet) supersymmetric case
and to $\frac{26-D}{3}$ in the ordinary bosonic string). Curiously
enough, the above metric (\ref{metrika}) with the dilaton field
{\em exactly} as in (\ref{dilatonio}) is a solution of the first
equation (\ref{uno}) for {\em any} value of $p$.
\par
The second equation, (\ref{dos}) however can be easily shown to be
given by:
\be
\frac{1-(p+1)/2}{(log u)^2} + \frac{1}{(log u)^2}
\ee
conveying the fact that it is a constant for $p+1 = 4$ only, in which
case it is zero. This means that the string must neccessarily be critical,
and we have to add the extra flat ({\em spectator}) dimensions, $21$
of them in the bosonic case \footnote{It should be clear that 
the spectator dimensions change
neither the confining string target space five-dimensional metric 
nor the Yang-Mills beta function. It is important to realize that
the metric (\ref{metrika}) does {\em not} correspond to a $D-$3 brane
in a $26$-dimensional ambient space. Actually, the only source for 
a nontrivial dilaton is the five-dimensional metric given by 
the running string tension.}. This is one remarkable fact of this background.
Another is, of course, that the {\em sign} of the dilaton in equation
(\ref{dilatonio}) is fixed, so that it is not possible to construct
{\em non-confinig} strings to describe ultraviolet slave quantum
 field theories using this set of ideas.
\par
The confining string itself should allow the computation of many
gauge invariant observables in gauge theories, such as the Wilson loop.
Assuming a static configuration bounded by $\sigma=\pm l/2$, and with
a temporal span equal to $T$, it is useful to work in the gauge in which
$\tau = x^0$ and $\sigma = x^1$. If at the same time we define a new coordinate
by $z\equiv log~ u$ such that the metric reads
\be
ds^2 = z d\vec{x}_4^2 + dz^2,
\ee
and assuming a symmetric imbedding of the world sheet into the target
spacetime of the confining string, $z=z(\sigma)$, then the induced metric 
on the worldsheet is given by:
\be
ds^2 = z d\tau^2 + (z+ z'^2) d\sigma^2
\ee
It can be argued (\cite{maldacena}) that the semiclassical 
approximation to the Wilson loop action is given by the area of the 
world sheet computed precisely with the induced metric, that is:
\be\label{w}
S = T \int_{-l/2}^{l/2} d\sigma\sqrt{z(z+z'^2)}
\ee 
(We are neglecting here the contribution of the dilaton to the action,
possibly important).
The variational principle (\ref{w}) enjoys a first integral, namely:
\be
p = \frac{- z^2}{\sqrt{z^2 + z z'^2}}
\ee
The formal solution to this differential equation is:
\be
\int^{z(\sigma)} df \frac{1}{\sqrt{f^3 - p^2 f}} 
= \int_{-l/2}^{\sigma}\frac{d\sigma}{p}
\ee
The lower limit of the first integral should be chosen in one of the regions
in which $z'^2$ is allowed to be positive by the first integral, namely
$(-p,0)$ or $(p,\infty)$. Choosing by concreteness it to be $0$, the solution
can be easily shown to be given by an elliptic integral of the first kind:
\be
\sigma + l/2 = - (2p)^{1/2} F(cos^{-1}(|z(\sigma)/p|),\frac{\sqrt{2}}{2})
\ee
which implies that:
\be
|\frac{z(\sigma)}{p}| = sn~ \frac{\sigma+l/2}{(2p)^{1/2}}
\ee
If we neglect the square of the modulus, $m^2 = 1/2$, the trigonometric
approximation can be applied:
\be
z(\sigma) = p sin~ \frac{\sigma + l/2}{\sqrt{2p}}
\ee
The relationship between $p$ and $l$ is found by demanding that
$z(l/2)=0$, yielding $p=\frac{l^2}{2\pi^2}$. The action then reduces to:
\be
S = T\int_{-l/2}^{l/2}d\sigma \frac{z^2}{p} = \frac{T l^3}{4}
\ee
which indeed corresponds to a confining potential.
\par
It would be interesting to check how world sheet supersymmetry (as well as
GSO projections) modify our analysis. In particular, it should be 
stressed that multiplicative factors in the beta function of the gauge
theory yield 
additive factors
in the formula (\ref{dilatonio}), which do not contribute to the string beta
function. 
\par
A quite surprising aspect of our analysis is the way pure Yang Mills 
perturbative information is promoted
 to string dynamics. This implies a connection between perturbative
aspects of gauge dynamics (such as the beta function) with non perturbative,
({\em stringy}) dynamics, such as infrared strong forces. The main
ingredient for doing that has been the soft dilaton theorem that allows 
to read coupling constant renormalization as variable
string tension and in that sense as confining string
background metric. It is perhaps worth remarking  that
vanishing $\sigma$-model beta functions - essential to
the quantum consistency of the string - have been used in
our approach only at the end as a consistency check
for the metric and dilaton field directly derived
from the quantum field theory data and not as an extra
governing principle. Moreover four dimensional
space time appears in this scheme as the only consistent
solution. All this seems to indicate that some
string dynamics is already encoded in the renormalization
of asymptotically free four dimensional gauge theories.

\section*{Acknowledgments}
This work ~~has been partially supported by the
European Union TMR program FMRX-CT96-0012 {\sl Integrability,
  Non-perturbative Effects, and Symmetry in Quantum Field Theory} and
by the Spanish grant AEN96-1655.  The work of E.A.~has also been
supported by the European Union TMR program ERBFMRX-CT96-0090 {\sl 
Beyond the Standard model} 
 and  the Spanish grant  AEN96-1664.
C.G would like to thank the organizers of the Abdus Salam ICTP
Extended Workshop in String Theory for hospitality
while this work was being written up.

\appendix


\end{document}